\documentstyle[aps,preprint]{revtex}

\newcommand{\lessim} {\mathop{\,<\kern - 1.05 em \lower 1.ex \hbox {$\sim$}\,}}
\newcommand{\grtsim} {\mathop {\,> \kern - 1.05 em \lower 1.ex \hbox {$\sim$}\,}}

\begin{document}
\normalsize
\draft
\title{ Theory for the Nonlinear Ultrafast Dynamics  of
Small Clusters after Ionization}

\author{M. E. Garcia$^1$, D. Reichardt$^2$ and K. H. Bennemann$^1$}

\address{$^1$Institut f\"ur Theoretische Physik der Freien 
Universit\"at Berlin, 
Arnimallee 14, 14195 Berlin, Germany.\\ 
$^2$Walther-Nernst Institut f\"ur Physikalische und Theoretische Chemie,
Humboldt Universit\"at zu Berlin, Bunsenstr. 1, 10117 Berlin, Germany}

\date{\today }
\maketitle


\begin{abstract}
The ultrafast relaxation of small clusters immediately after
ionization is studied.  For small Hg$_n$ clusters we determine 
 the 
fragmentation-time distributions  induced by 
ionization. 
 A dramatic change of the fragmentation behaviour  occurs 
 when the temperature before ionization reaches the 
melting point of the neutral clusters. 
 The ultrafast dynamics depends nonlinearly on initial conditions
regarding atomic positions and velocities. The resultant largest 
Lyapunov exponent increases strongly upon ionization.  
\end{abstract}



\newpage

A fundamental problem in the physics of small clusters and molecules is
 the description of the relaxation mechanisms in the sub-picosecond
time-domain. The ultrafast dynamics (UD) of a cluster is usually induced  
  with the help of an ultrashort laser pulse which
excites or ionizes the cluster, abruptly transferring it to a nonequilibrium
state. It is presently still
unclear how the excess energy released by this process 
is transferred  to the various modes of the cluster.  

 Since the recent development of the Femtosecond 
Spectroscopy\cite{zewail}, which permits a time-resolved monitoring 
of the  atomic motion and fragmentation processes, many different  
pump-probe experiments have been performed to study the UD 
of excited and ionized clusters\cite{woeste-manz,gerber,nenepo}.  
 The experimental results are usually interpreted in terms of
master-equations, assuming constant decay probabilities. This
 allows to obtain some information about
 decay-times due to  fragmentation processes. 
 However, many questions still remain open. For instance, not much is 
 known about the time-scales 
 for energy transfer from the electronic to the atomic degrees of
freedom. 
 Moreover, one of the 
most important 
problems which has not even been addressed so far, is the dependence of the
 UD on the initial conditions, and, in particular, on the
initial temperature and cluster size. A temperature-sensitive   
ionization-induced short-time dynamics 
could  allow to use the short-time spectroscopy after ionization as a 
method  to determine the temperature of the clusters before ionization. 
 Of particular interest is also whether the fragmentation of clusters
 is sensitive to phase transitions like the solid-liquid transition and
if it reflects the intrinsic nonlinear, chaotic behaviour 
of clusters\cite{berry,indu}.

The ionization induced changes in
the electronic structure  are particularly strong in van der
Waals systems, like  small
Hg$_n$ clusters\cite{haber,mgb}, which leads to a rapid fragmentation, as
it has been observed in recent pump-probe experiments\cite{gerber}.

It is the purpose of this letter to study the dynamics of small Hg$_n$ clusters as 
prototype for strong
response to ionization. Of course,  for clusters
with other bond character the relaxation time 
scales are expected to be different. However, the
general  features of ionization induced fragmentation and underlying
mechanisms of van der Waals clusters should be of general validity. 

 We describe the UD of small Hg$_n$ clusters immediately after 
 ionization with the help of electronic structure calculations and 
molecular dynamics (MD) simulations in the Born-Oppenheimer
approximation. 
 Usually, for MD simulations of  neutral  van der Waals
clusters, parameterized pair potentials are used\cite{jellinek}.
   However, this description is no longer valid when these clusters
become ionized. 
  The electronic ground state of
ionized van der Waals clusters is governed by the interplay between
delocalization energy of the positive charge- and polarization
energy\cite{haber,mgb}. Thus, in order to account for the
time-dependent changes of the charge- and dipole distributions during the 
  relaxation following ionization, 
 the potential energy surface (PES) must be determined self-consistently.  
 Therefore, we use a 
 microscopic electronic theory based on the ionic-core
model\cite{haber}. 
  We assume that the hole created 
in the ionization process delocalizes only within a dimer. 
 Hence, after  imposing this constraint for the charge distribution,
 and 
 neglecting charge-dipole fluctuations, 
   the ground-state energy of the ionized cluster is given by
\begin{eqnarray}
\label{energy}
 E &=& \frac{(\phi_1 + \phi_2)}{2} - \sqrt{t_{ss}^2({\bf r}_{12})
 + \frac{(\phi_1 - \phi_2)^2}{4}}  \nonumber \\ 
 &&+ \; \varepsilon \sum_{k,k'\atop
 k\not= k'} \biggl[ \; \biggl(\frac{r_0}{r_{k k'}} \biggr) ^{12} - 2 \biggl(
 \frac{r_0}{r_{k k'}} \biggr) ^6 \; \Biggr]
 \; \nonumber \\
 &&- \; \frac{1}{2}\;  \alpha  \!\! \sum_{k \in \{n-2\}}
 \mid {\bf \cal E}_k^Q + {\bf \cal E}_k^P \mid ^2 + \;\;E_{dc}(\{\langle {\bf
P}_k \rangle \}, \{ \langle Q_l \rangle\}). 
\end{eqnarray}
Here, the first two terms  refer to the energy of the ionic dimer, where
$t_{ss}$ is the interatomic $ss$ hopping element  and
 $\phi_l$ is the potential  of the induced dipole 
 distribution   at the site  $l$  of the dimer core.
   The third term describes the van
     der Waals energy and the core-core repulsions in the whole cluster,
being $\varepsilon$ and $r_0$ the cohesive energy and the bond-length of 
Hg$_2$, respectively.  
       The fourth term refers to the energy of the
         neutral subcluster. 
           $\alpha$ is the atomic polarizability, whereas
${\bf \cal E}_k^Q$
  and ${\bf \cal E}_k^P$ are the charge- and dipole fields on the
neutral atom
    $k$. Finally, $E_{dc}$  
       stands for the double-counting energy arising from the decoupling
        procedure used\cite{mgb}.

   The expectation values of the charge- ($\langle Q_l \rangle$) and 
        dipole distributions ($\langle {\bf P}_k \rangle$)
           are coupled through the nonlinear relations 
                       $\langle {\bf P}_k \rangle = \alpha \; ({\bf  {\bf
\cal E}}_k^Q + {\bf {\bf \cal E}}_k^P)$ and     $\langle Q_l 
\rangle = |e| \sum_{\sigma} 
 \langle n^h_{l s \sigma} \rangle$ (where $\sum_{\sigma} \langle 
 n^h_{l s \sigma} \rangle$ is the 
number of holes at atom $l$)\cite{mgb}. 
 The PES for the motion of the atoms is thus
defined by the function  $E({\bf r}_1, ..,{\bf r}_n)$. 
 Consequently, the $\mu$ component of the force acting on a given atom $i$ is given by
$F_i^{\mu} = - \partial E/\partial {\bf r}^{\mu}_i$. 
 Note, that the energy $E({\bf r}_1, ..,{\bf r}_n)$ cannot be described
as a sum of pair potentials. 

 Assuming that the width of the ionizing laser pulse is negligible compared 
 to  the time scale of the nuclear motion,  the main physics of the
dynamics described by this theory can be visualized as follows. 
 Through the
 ionization process, the Lennard-Jones PES of the neutral cluster 
  is modified by two rather strong 
 attractive terms, hole delocalization- and charge-dipole energy, 
 and one mainly repulsive but weaker term, 
 the dipole-dipole interaction energy.
 The binding energy of the ionized cluster is much larger than that of
 the neutral cluster, and equilibrium interatomic distances are considerably 
 smaller. 
 Thus,  immediately after the vertical ionization, a relaxation process 
 follows, for which the anharmonic part
 of the PES plays an important role.  
   The  excess energy $\delta
   E$ comes from
   the charge-dipole interactions and from the delocalization energy of the
   hole. The latter contribution is the largest, and therefore for
increasing times there will  be a net transfer
of kinetic energy from the dimer to the neutral rest of the cluster. 

 Using this theory we determine the fragmentation behaviour and  its
dependence on initial temperature for Hg$_n$ with $ 3 \le n \le 13$. 
 For the parameters 
    $\varepsilon$ and $r_0$ we use experimental values\cite{vanzee},
whereas 
         $\alpha$ is obtained from atomic
            data\cite{gummi}. For the distance dependence of the 
            hopping element $ss$
               we use the functional form proposed in Ref.~ 
                  \cite{petifor}.
  To integrate the equations of motion we use the
    Verlet algorithm in velocity form, and calculate the forces through
the above described 
self-consistent
  procedure for each time step $\Delta t$ ($\Delta t = 7\cdot \; 10^{-16}s$).
  As a starting point, we determine the atomic structure of the
neutral clusters by simulated annealing. 
 Then, by using standard numerical procedures\cite{jellinek} we generate, for 
 each size, a distribution of clusters at a
given temperature $T$, where $T$ is defined as the time-average of the kinetic 
energy over a long MD run ($\sim 10^6$ time-steps). 
  This ensemble of clusters is ionized, i.e., the PES 
  is switched from the neutral to the ionized
electronic state.  The dimer core is assumed to be formed by the pair of
atoms closest to each other.

 We obtain ionization induced fragmentation for all clusters
sizes under consideration. 
 The neutral fragments are mainly   
  monomers, but dimer, and even trimer emission
can also occur. In order to have a rough estimation of the fragmentation
times, we performe first calculations at zero initial temperature and
highly symmetric initial structures. 
Results  show for very 
small clusters ($3 \lessim n \lessim 6$) a fast emission of the first 
 fragment, which results from a rapid energy transfer into particular
modes of the cluster ("cold" fragmentation).
 The corresponding fragmentation time $\tau_F$, i.e.,  the time at which the 
 first fragment leaves the original cluster, is rather independent of the
cluster size, and  smaller than $10 {\rm ps}$.    
In contrast, for $n = 13$,  $\tau_F$ is larger than $30 {\rm ps}$,
and  there are no longer privileged modes to which the excess energy is 
transferred. As a consequence, some  thermalization takes place before 
evaporation of atoms.  
 Our results for $\tau_F$ for the different sizes at $T= 0$ are in 
 relative good agreement with 
 those determined from pump-probe experiments\cite{gerber}.  However, 
 in order to simulate real experimental conditions,   
  clusters must be given a 
nonzero initial temperature. 
 For this purpose we study the fragmentation behaviour 
  on a distribution of approximately 1000 clusters for each 
initial temperature.   
 Thus, we obtain a  distribution of fragmentation-times ${\cal N} (\tau_F)$,
defined by the number of clusters whose first fragmentation occurs at
$\tau_F$. 

In Fig.~1 (a) we show the normalized fragmentation-time
distribution (FTD), $W(\tau_F)$,
calculated  for Hg$_3$ clusters with an initial temperature of $T = 40
K$. Note that  one can interpret the function $W(t)$ as the probability
 of fragmentation as a function of time.  It becomes clear,
that for $T\not= 0$ one cannot distinguish anymore between cold and
thermal fragmentation as a function of the cluster size. For all sizes
studied, both forms of fragmentation are present. The FTD of Fig.~1 (a)  
  exhibits  two
general features common to all calculated distributions (s. also Fig.~1(c)). 
 It has a
certain width (usually larger than $10 {\rm ps}$), 
 and its weight at short times is small. It is important to
remark that the calculated decay-probability $W(t)$ is not constant, as
is usually assumed. 
 In contrast to what one would expect,  
  the width of $W(t)$ does not decrease for
decreasing temperature. 
 Since lower temperature implies smaller displacements of the atoms
around their equilibrium positions,  
 this is a clear indication of a nonlinear dependence of $\tau_F$ on the
initial conditions. 
 For the sake of illustrating this effect, 
 we determine the FTD for 
 Hg$_3$ clusters at an extremely low temperature, for which the atomic
displacements are of the order of $10^{-6}{\rm \AA}$\cite{nonli}.  

 In Fig~1 (b) the resulting FTD is shown. The salient feature is a  
 still large width,   
   the
presence of a gap and several sharp  peaks. 
 This strange form is another indication of the 
intrinsic nonlinear behaviour  of the fragmentation dynamics. Small
clusters are chaotic systems\cite{berry,indu}. This means that the 
trajectories in phase
space are very sensitive to the initial conditions. Since
the fragmentation times strongly depend on the trajectories of the
ionized clusters on the PES, it is clear that they should exhibit
nonlinear behaviour.  The trajectories corresponding to the various 
peaks  of $W(t)$   in Fig~1 (b) 
 could be somehow 
interpreted as "attractors", since the resulting fragmentation 
times occur 
starting from many
different initial configurations.   
 We obtain this chaotic behaviour after fragmentation for    all
cluster sizes studied ($n > 2$). A very small change of the initial
conditions  leads to completely different trajectories which involve
different processes (e.g., cold- or thermal- or even no fragmentation). 
 Note that in our simulations energy is conserved up to $10^{-4} 
{\rm eV}$ for $10^5$ time steps after ionization.  

 Of course, at very low temperatures quantum effects must dominate and
will tend to 
smear out the nonlinear features. But at low temperatures above the zero
point motion, where the mean displacements of the atoms are still very
small ($\sim 0.05 {\rm \AA}$) the chaotic behaviour might play a role in
the fragmentation dynamics.

 In the inset of Fig.~1(b)  we show the time evolution of the
total polarization of one of the clusters which contributes to the first
peak. Notice that, even in the case of fastest fragmentation, the neutral atom
needs a few oscillations against the charged dimer in order to gather 
the kinetic energy necessary
to escape. This  arises from the fact that the frequencies of this mode
and the inner vibrational mode of the dimer core are different (as is
reflected in the time-dependence of the polarization) and consequently  
 an
efficient energy transfer is prevented. At high  temperatures, however, 
there is a
nonzero probability for fragmentation within the first oscillation
period [s. Fig.~1(a)].  
 When the neutral atom leaves the cluster, its induced 
dipole moment starts
to decrease with time and goes to zero as the distance to the ionic
dimer  becomes large.

We have already pointed out that all our calculated distribution functions
$W(t)$ exhibit a very small weight at short times. This means that the
probability for fragmentation is small at times shorter than the
characteristic times for excess-energy transfer. If one writes a master
equation for the decay of the original cluster, one obtains for the 
time-dependent number of clusters with the original size $n$ 
\begin{equation}
N_n(t) = N_n(0) e^{-\int_0^t dt' W(t')},   
\end{equation}
 and $d^2 N_n / dt^2 \mid_{t=0} < 0$ 
 in agreement with what 
 is observed in pump probe experiments\cite{woeste-manz,gerber}, and in
contrast with what one would obtain by assuming a constant $W$. 

In general, our calculated FTD show only a week temperature
dependence. The same holds for the corresponding average fragmentation
times $\langle \tau_F \rangle$. But as soon as the temperature 
 before ionization crosses the solid-like to liquid-like  transition
temperature of the neutral clusters 
$W(t)$ undergoes strong changes. This indicates a correlation between the 
energy
transfer mechanisms after ionization and the thermodynamical state of 
the cluster
before ionization.   In Fig.~2 (a), (b) and (c) we show the temperature
dependence of of the root-mean-square (rms) bond-length fluctuation $\delta$,
defined as usual as 
$ \delta = (n(n-1)/2)^{-1} \sum_{i<j} \sqrt{(\langle {\bf r}^2_{ij} 
\rangle - 
\langle {\bf r}_{ij} \rangle^2 )}/\langle r_{ij} \rangle$,  
  where $\langle ... \rangle $ means time-average over a long trajectory
($10^6$ time steps), which gives an idea of the degree of mobility of
the atoms. 
 For bulk material, as a function of the temperature, $\delta$ shows 
 typically  a sharp 
increase at the solid-liquid transition, consistent with the
Lindemann-criterion. This behaviour has been used as a numerical tool 
 to determine the
melting-temperature of small clusters\cite{jellinek}. From Figs.~2 (a), 
(b) and (c) the temperature dependence of $\delta$ is shown for $n =
3,4$ and $13$. 
 For $n = 3$ and 13, $\delta$ clearly shows a jump. The smoother increase 
 of $\delta$ 
for $n = 4$ is related to the interplay (isomerization) between the 
tetrahedral and the planar rombic structure, which occurs at a
temperature lower than the actual melting point and gives rise to larger
values of $\delta$, smearing out the sharp increase at $T_M$. 
Now, the  important
result of our calculations becomes obvious by noting that the inverse of
the average fragmentation times $\langle \tau_F \rangle$ shows {\it
exactly the same} temperature dependence as $\delta$. This occurs even
for $n = 4$, reflecting a strong sensitivity of $\langle \tau_F \rangle$
to the melting dynamics. 
 This result can be interpreted as follows. 
  The distribution of the excess energy $\delta E$ 
 among the different degrees of freedom leads (except for the
unlike cases
of cold fragmentation) to an homogeneous
weakening of the bonds. 
 Only those 
atoms in a liquid-like environment have a larger probability to  
evaporate. If 
  the neutral cluster is already liquid, fragmentation can occur
faster, since $\delta E$ can be fully used for evaporation. 
 Note that the above  argument is  independent of the bond character of
the clusters considered. However, it is essential that a
 vertical ionization puts the cluster in a  non-equilibrium situation.
For an adiabatic ionization the above picture does not necessarily hold, 
because the
melting temperature of the ionized cluster could be much larger than
that of neutral one. 
 Obviously, 
if $\delta E$ is smaller than the 
latent heat of the cluster, fragmentation may  only occur above the
melting point.  
 We
believe that this effect can also be observed in metallic and covalent
clusters,  as long as the conditions discussed above are fulfilled.

The remarkable correlation between  $\langle \tau_F \rangle$ and 
$\delta$ could allow to determine the solid-like to liquid-like  transition 
of small
clusters by pump\&probe experiments. 
 From the experimental signal $N_n(t)$ 
  one can 
 first obtain $W(t)$ by using
Eq.~(2), and then calculate 
    $\langle \tau_F
\rangle = \int_0^{\infty} dt\, t \, W(t)$. 

In order demonstrate the importance of chaos in the short-time dynamics
and to quantify the influence of ionization in the chaotic
behaviour of clusters, we determine the corresponding local 
Lyapunov exponents\cite{berry} before and
after
ionization process. The largest Lyapunov exponent is a measure for the 
average divergence of two initially nearby
trajectories in the phase space, and is  given by 
\begin{equation}
  \lambda = \lim_{N\to \infty} \frac{1}{N} \;
log \mid 
J(\{{\bf r}\}_n,\{{\bf p}\}_n)\mid,
\end{equation} 
where  $J(\{{\bf r}\}_n,\{{\bf p}\}_n)$ is the Jacobian matrix for the map from
configuration at time $t$ to the configuration at time $t + \Delta t$ in
the Verlet algorithm\cite{berry}.  We obtain positive values of 
$\lambda$ (i.e., chaotic behaviour) 
for all cluster sizes studied. 
In Fig~3 we show the temperature dependence of the (averaged) largest 
Lyapunov exponent  for Hg$_4$ clusters
before ($\lambda^0$) and after ($\lambda^+$)  ionization. At a 
given initial temperature, the calculated $\lambda^+$ 
 are more
than one order of magnitude larger  than $\lambda^0$. 
This clearly indicates that degree of nonlinearity  is enhanced by the
ionization process. 
 Due to the large  energy $\delta E$ pumped into the atomic
degrees of freedom, the cluster can explore a larger part of the PES,
which lead to an increase of the nonlinear behaviour. 
 Since $\delta E$  is of the order of $10^4 K$ it is also clear that
$\lambda^+$ is almost independent of the initial temperature. 
 This explains why the width of the FTD does not go to zero for $T \to
0$ [s. Fig.~1(b)]. 
In contrast, $\lambda^0$ exhibits an appreciable  increase exactly
within the temperature range where the solid-liquid transition occurs. 
 This interesting behaviour of $\lambda^0$ (which has been also observed
for Ar$_n$ clusters\cite{indu}) reflects the fact that in the liquid  phase the
motion is more chaotic than in the solid phase.

This work has been supported by the Deutsche
Forschungsgemeinschaft through the SFB 337.


\newpage

\begin{figure}
\caption{Calculated ionization induced fragmentation-time distributions $W(t)$
 for a)
Hg$_3$ clusters at initially $T = 40K$, b) Hg$_3$ clusters at initially
$T = 0.05K$, and (c)Hg$_6$ clusters with $T = 40K$. 
Note that the curves have been normalized so that $W(t)$ also
represents the probability of fragmentation at a time $t$. Inset of
Figure (b): time behaviour of the polarization of a cluster which
fragmentates at $t = \tau_1$.}
\label{Figure 1}
\end{figure}

\begin{figure}
\caption{Temperature dependence of the inverse average fragmentation
times $\langle \tau_F \rangle^{-1}$ of the ionized 
clusters (up triangles, left axis) and rms bond-length fluctuations
(open circles, right
axis) $\delta$ before ionization for a) Hg$_3$, b) Hg$_4$ and 
c) Hg$_{13}$ clusters.
 Note that the increase in $\langle \tau_F\rangle^{-1}$ 
characterizes the melting
temperature,  and could be used for experimental determination 
of the solid-liquid transition. }
\label{Figure 2}
\end{figure}

\begin{figure}
\caption{Calculated temperature dependence of the largest local Lyapunov 
 exponent $\lambda(T)$ for Hg$_4$ clusters
 before ($\lambda^0(T)$) and after ($\lambda^+(T)$) ionization. Note 
 that the ionization process enhances the chaotic behaviour of 
 the clusters, by more than one order of magnitude. 
 The increase of $\lambda^0(T)$ for increasing $T$ reflects the
solid-liquid transition.}
\label{Figure 3}
\end{figure}


\begin{references}
\bibitem{nonli} Note that
this corresponds to a kinetic energy smaller than that of the zero point motion 
of Hg$_3$ clusters
and cannot be used to compare with experiment.
\end{references}
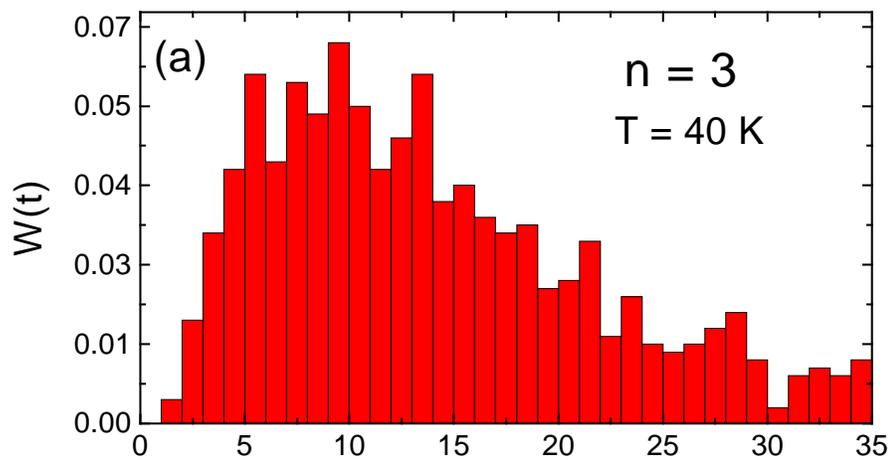
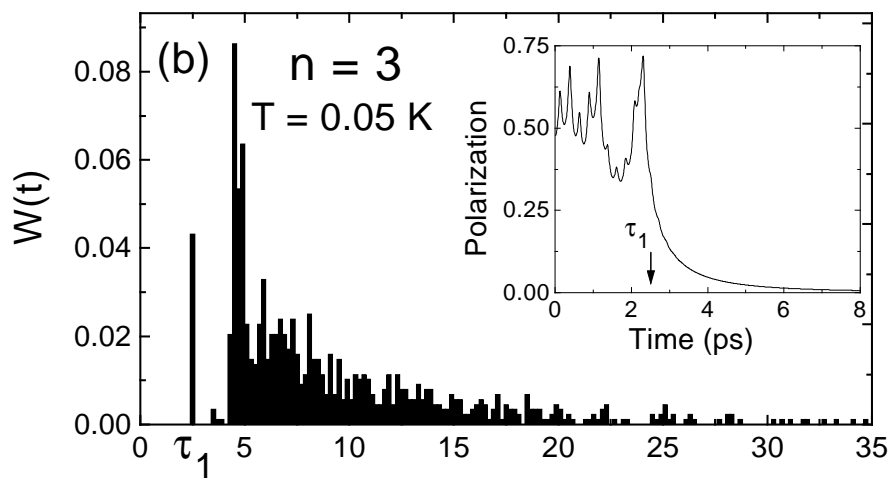
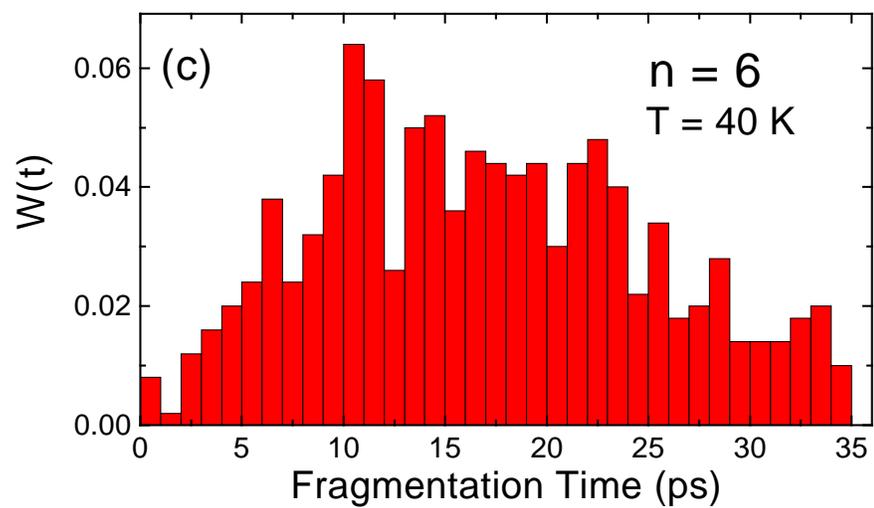
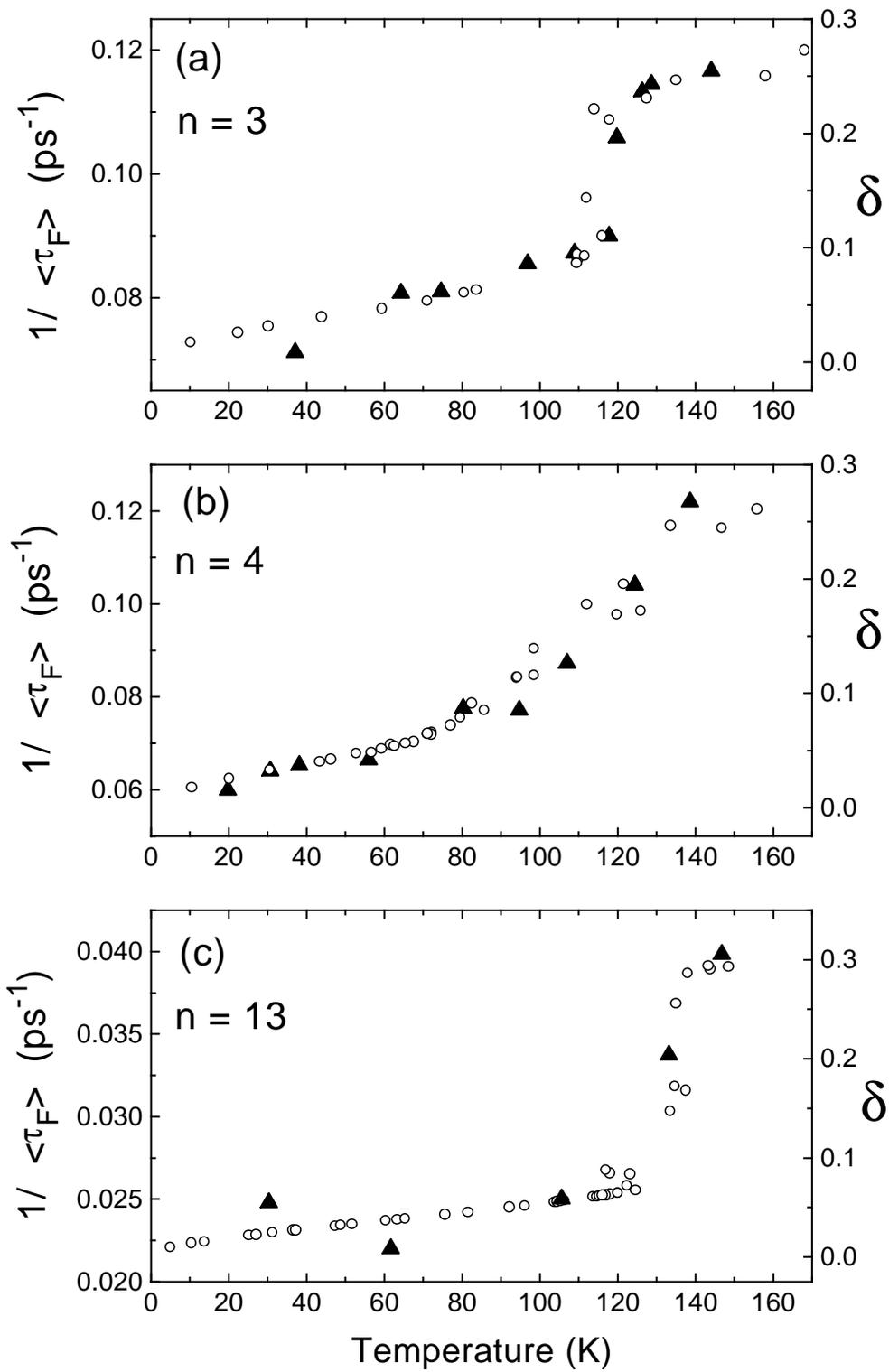
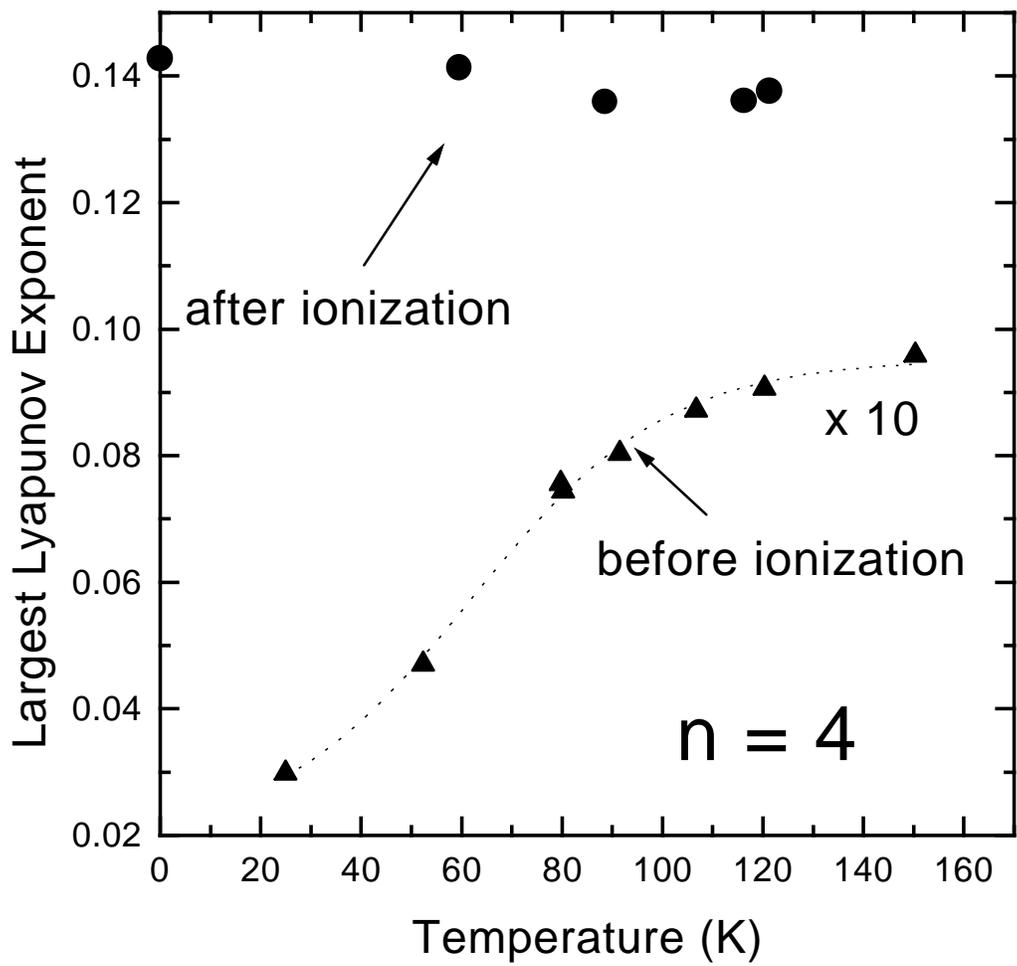
\end{document}